\documentclass{iau} 
\usepackage{graphicx}

\title[The dark energy of CMEs] %% give here short title %%
{Stellar flares and the dark energy of CMEs}

\author[J.J.~Drake et al.]   %% give here short author list %%
{Jeremy J.~Drake$^1$, Ofer Cohen$^1$,
 Cecilia Garraffo$^1$ \and  V.~Kashyap}
\affiliation{$^1$Smithsonian Astrophysical Observatory, 60 Garden Street, Cambridge MA02138, USA\\email: {\tt jdrake@cfa.harvard.edu}}

\pubyear{2016}
\volume{320}  %% insert here IAU Symposium No.
\jname{Solar and Stellar Flares and Their Effects on Planets}
\editors{A.G. Kosovichev, S.L. Hawley \& P. Heinzel, eds.}
\begin{document}

\maketitle

\begin{abstract}
Flares we observe on stars in white light, UV or soft X-rays are
probably harbingers of coronal mass ejections (CMEs).  If we use the
Sun as a guide, large stellar flares will dissipate two orders of magnitude less X-ray radiative energy than the kinetic energy in the associated CME.  Since coronal emission on active stars appears to be dominated by flare activity, CMEs pose a quandary for understanding the fraction of their energy budget stars
can spend on magnetic activity.  One answer is magnetic  suppression of CMEs, in which the strong large-scale fields of active stars entrap and prevent CMEs unless their free energy exceeds a critical value.  The CME-less flaring active region NOAA 2192 presents a possible solar analogue of this. Monster CMEs will still exist, and have the potential to ravage planetary atmospheres.
\end{abstract}

\firstsection % if your document starts with a section,
%              \vspace{-0.2in}

\section{Introduction}

In the last decade or so, it has become clear that the energy expended in large flares on the Sun is dominated not by the thermal radiation in the X-ray and EUV regimes that we have become so accustomed to studying, but by the energy of the  coronal mass ejections (CMEs) that are associated with the magnetic field re-organization, followed closely by the UV-optical ``white light'' continuum.   Studies of the timing and location of CMEs have revealed that the fraction of flares accompanied by CMEs increases with flare energy, until X-class  (peak GOES 1--8~\AA\ flux $> 10^{-4}$~W~m$^{-2}$ as seen at Earth) when essentially all flares are accompanied by a CME (e.g., \cite{Yashiro.Gopalswamy:09}; see Figure~\ref{f:evlac}).

  At 4.6 billion years old, the Sun is a fairly quiet, inactive star compared with its magnetic behavior at younger ages.  In the T~Tauri to zero-age main-sequence phase, stars like the Sun rotate  ten times faster and can be 1000--10,000 times more active in terms of X-ray output (see, e.g., the discussion of \cite[Wright et al.\ 2011]{Wright.etal:11}).  Magnetically-active stars exhibit what would be absolute monster flares if they were to occur on the Sun.  Large flares on M dwarfs can readily radiate $10^{34}$~erg in soft X-rays, while the largest flares observed, on active RS~CVn and Algol-type binaries, reach X-ray fluences of  $10^{37}$~erg (e.g., \cite[Favata 2002]{Favata:02}).  In contrast, the largest GOES flares reach a total fluence of $\sim 10^{31}$~erg (or about twice this value in the 0.1-10~keV range).
  %\footnote{Equivalent to approximately twice this value in the 0.1-10~keV range.}

As we have been frequently reminded during this meeting, solar flares can have a dramatic affect on the Earth's space radiation environment, causing aurorae and geomagnetic storms that have been known to disable satellites and terrestrial electric power grids.  In this context, the obvious question is what is the analogous impact of much larger stellar flares on their planetary progeny?  CMEs are extremely difficult to detect on other stars with current instrumentation and no definitive observations exist for stars other than the Sun. We have no option but to look to the Sun for clues as to how CMEs might scale to stellar flare energies.  
%Some key questions arise:
%\begin{enumerate}
%\item Can we really extrapolate from the Sun?
%\item Are close-in planets around active stars doomed by giant CMEs?
%\item What are stellar CME mass and angular momentum loss rates?
%
%As we shall discover, implied CME energy losses are substantial, which raise one further %question:
%
%\item What fraction of Lbol can dynamo activity extract?
%\end{enumerate}

%Current hopes lie in radio detections of the ``Type II burst" 

\vspace{-0.2in}

\section{If stellar CMEs were direct analogs of solar ones}

\begin{figure}
% \vspace*{-2.0 cm}
\begin{center}
\includegraphics[width=3.2in]{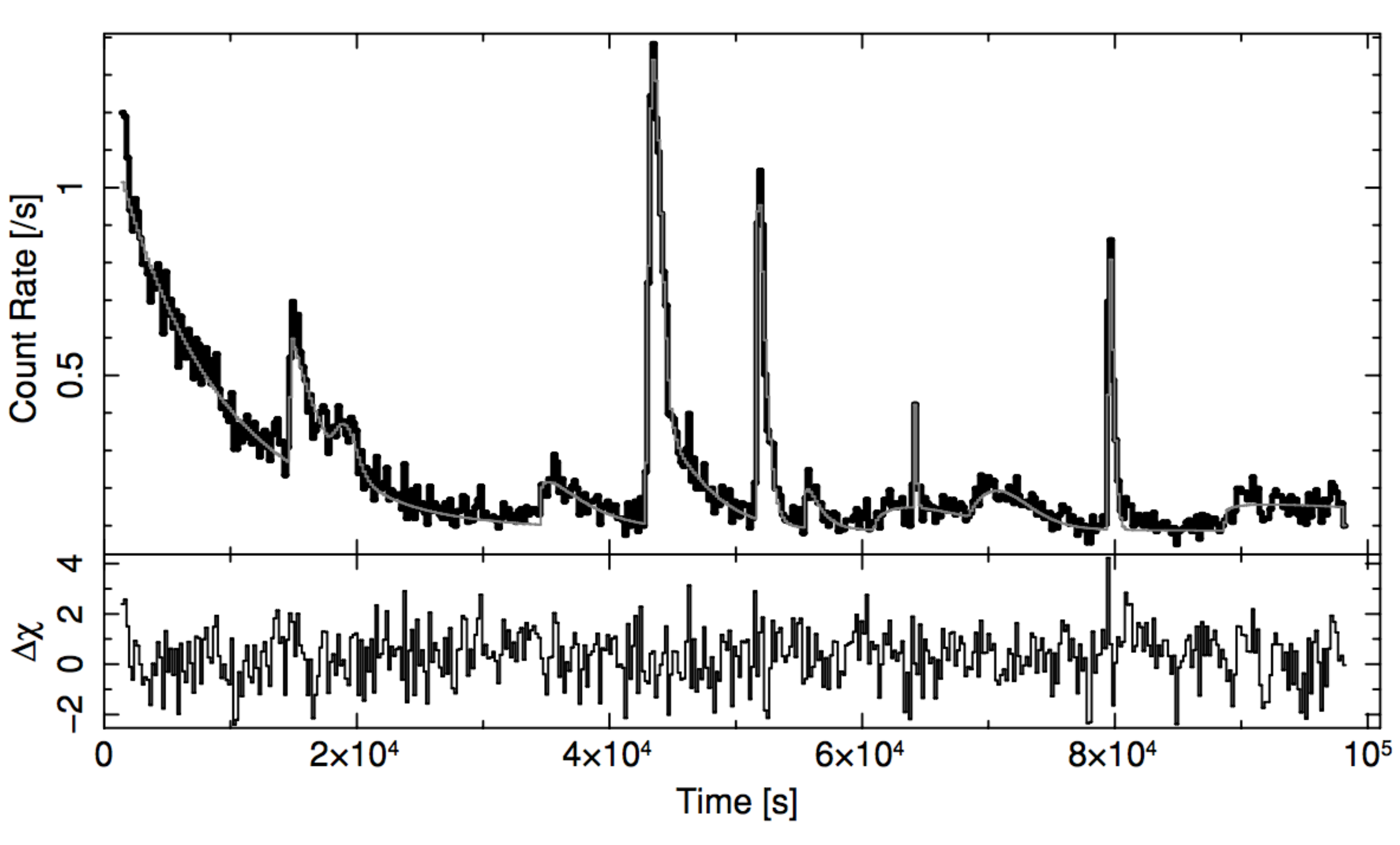}
\includegraphics[width=2.0in]{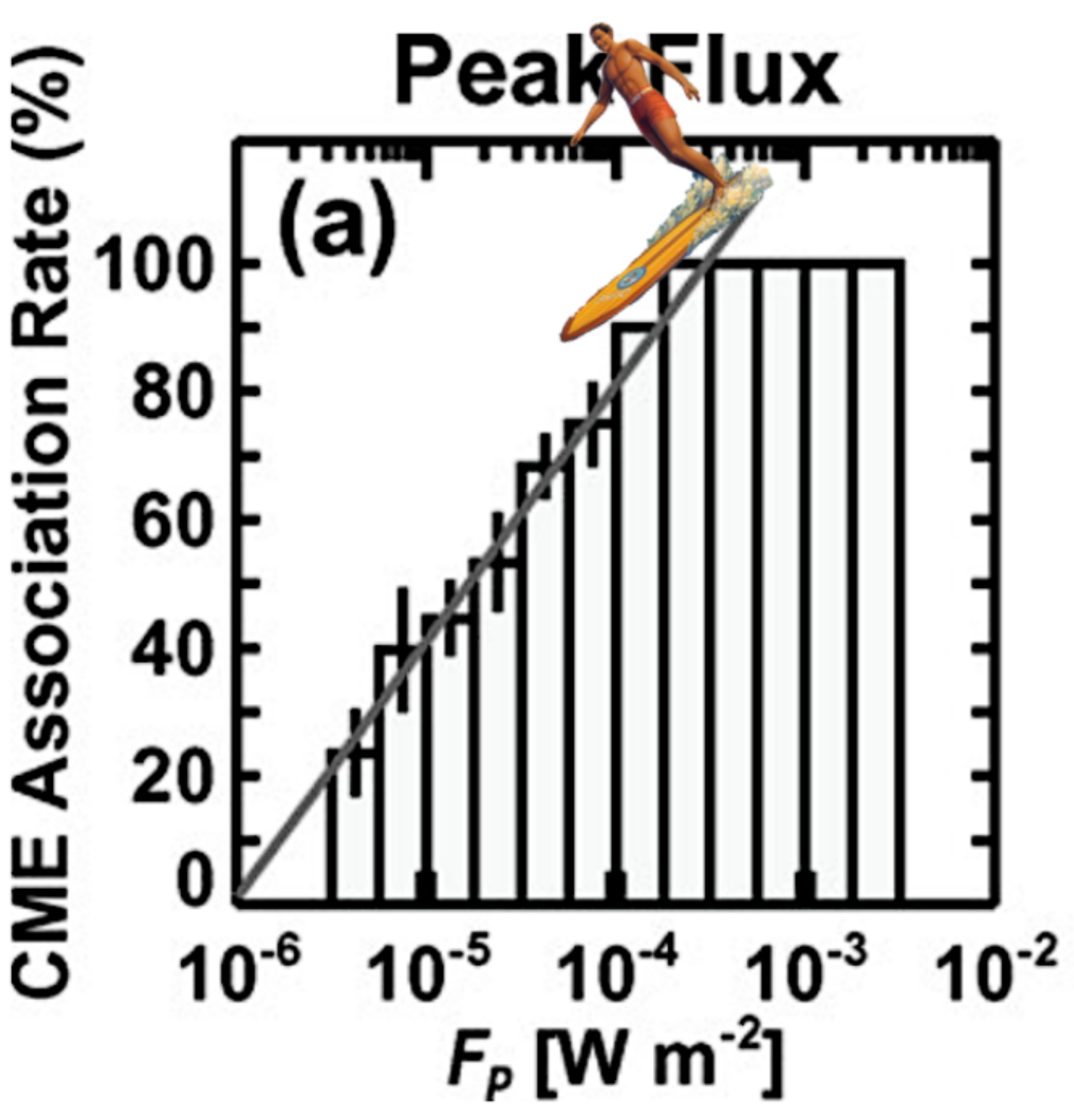} 
% \vspace*{-1.0 cm}
\caption{Left: The 1--25~\AA\ X-ray light curve of the dM3.5e star EV~Lac from a 2009 March 13 {\it Chandra} HETG observation (ObsID 10679;\cite[Huenemoerder et al.\ 2010]{Huenemoerder.etal:10}) compared with an analytical model fit. Right: The solar CME-flare association fraction from \cite[Yashiro \& Gopalswamy (2009)]{Yashiro.Gopalswamy:09}}
 \label{f:evlac}
\end{center}
\end{figure}

Close examination of the EUV and X-ray light curves of active stars indicates that most of the flux appears to be in the form of flares---some large that we can readily see, but most much smaller so that they form a quasi-continuum of coronal emission (e.g., \cite[Audard et al.\ 2000, Kashyap et al.\ 2002]{Audard.etal:00,Kashyap.etal:02}).  An X-ray light curve of the dM3.5e star EV~Lac is shown in the left panel of Figure~\ref{f:evlac}.  Several flares of different intensities are readily visible, sitting on a pseudo-continuum.  The indication is that all the wrinkles and crinkly bits at smaller amplitudes are actually multiple flares of lower intensity. 

\begin{figure}
% \vspace*{-2.0 cm}
\begin{center}
\includegraphics[width=2.9in]{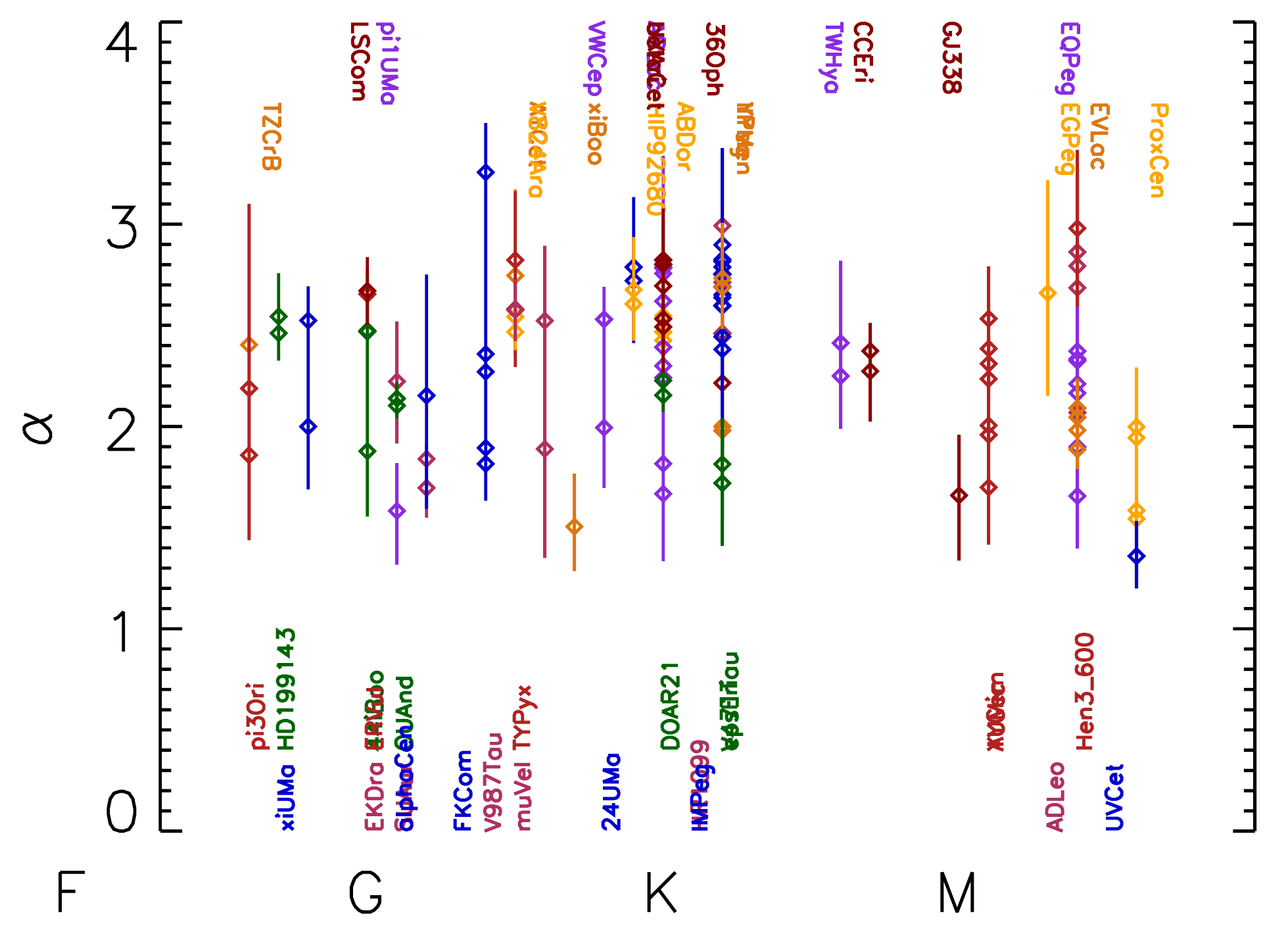}
\includegraphics[width=1.7in]{duke_orig.png}
% \vspace*{-1.0 cm}
\caption{The flare frequency as a function of energy power law index $\alpha$ derived from photon event arrival times in {\it Chandra} observations of different stars (from Kashyap et al., in prep.).}
 \label{f:alpha}
\end{center}
\vspace{-0.1in}
\end{figure}

As has been discussed extensively in the solar context, the flares tend to follow a power law distribution in frequency as a function of flare energy of the form
\begin{equation}
%\hspace{-1.0in}
%\includegraphics[height=0.2in]{duke_orig.png}
%\hspace{1.0in}
\frac{dn}{dE}=k E^{-\alpha}\, \;\;\;\includegraphics[height=0.2in]{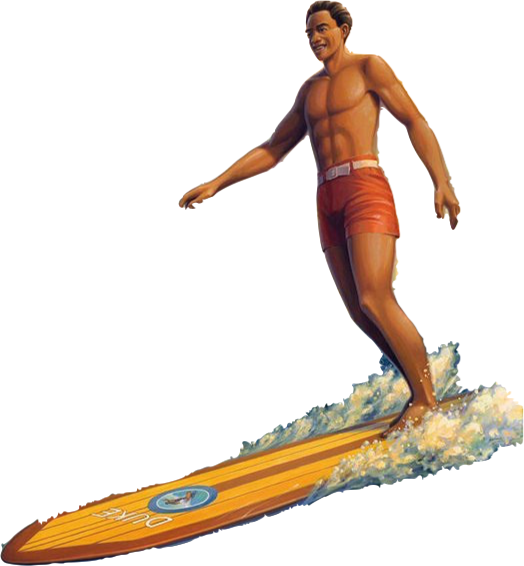}\hspace{0.3in}
 k=\frac{L_X (2-\alpha)}{\left(E_{max}^{2-\alpha} - E_{min}^{2-\alpha}\right)},
\label{e:hudson}
\end{equation}
where $k$ is a normalization constant that can be obtained by equating the flare energy, $E$, to its X-ray fluence.  For the Sun, values of $\alpha$ between 1.5--2.5 have been found, depending somewhat on the wavelength range of study.  For flare-dominated coronal emission, we can also equate the integration of Eqn.~\ref{e:hudson} over energy to the observed stellar X-ray luminosity.  Figure~\ref{f:alpha} shows the results of an analysis of X-ray photon arrival times in a large number of {\it Chandra} observations of late-type stars that indicates values of $\alpha$ are in the range 1.5--3 or so.  These numbers are significant in that, for $\alpha\geq 2$, the integral tends to infinity as the lower limit of integration (the lowest flare energy considered) tends to zero---physically speaking, the flares can indeed explain all the emission.

\begin{figure}[b]
\includegraphics[width=2.6in]{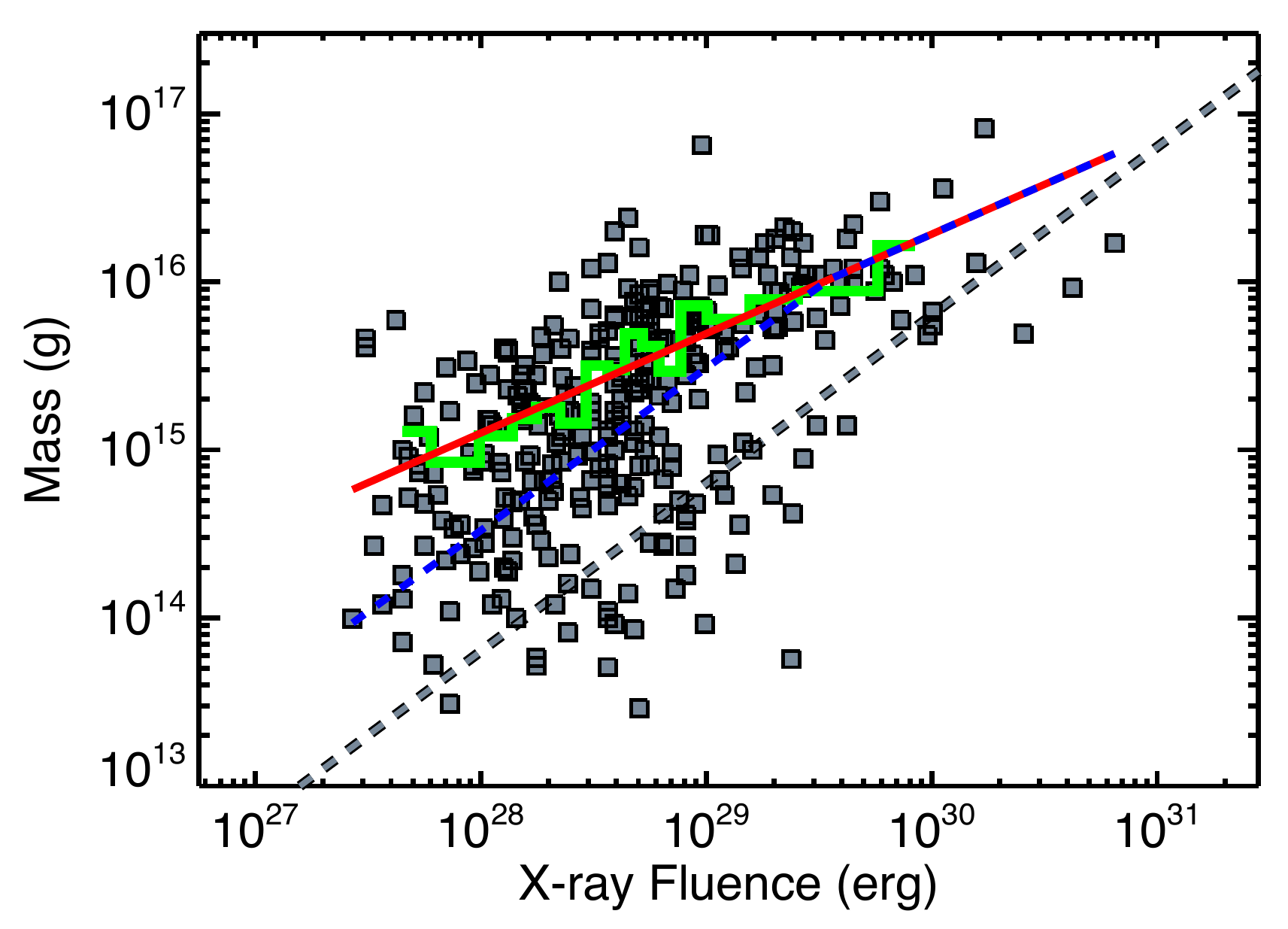}
\includegraphics[width=2.6in]{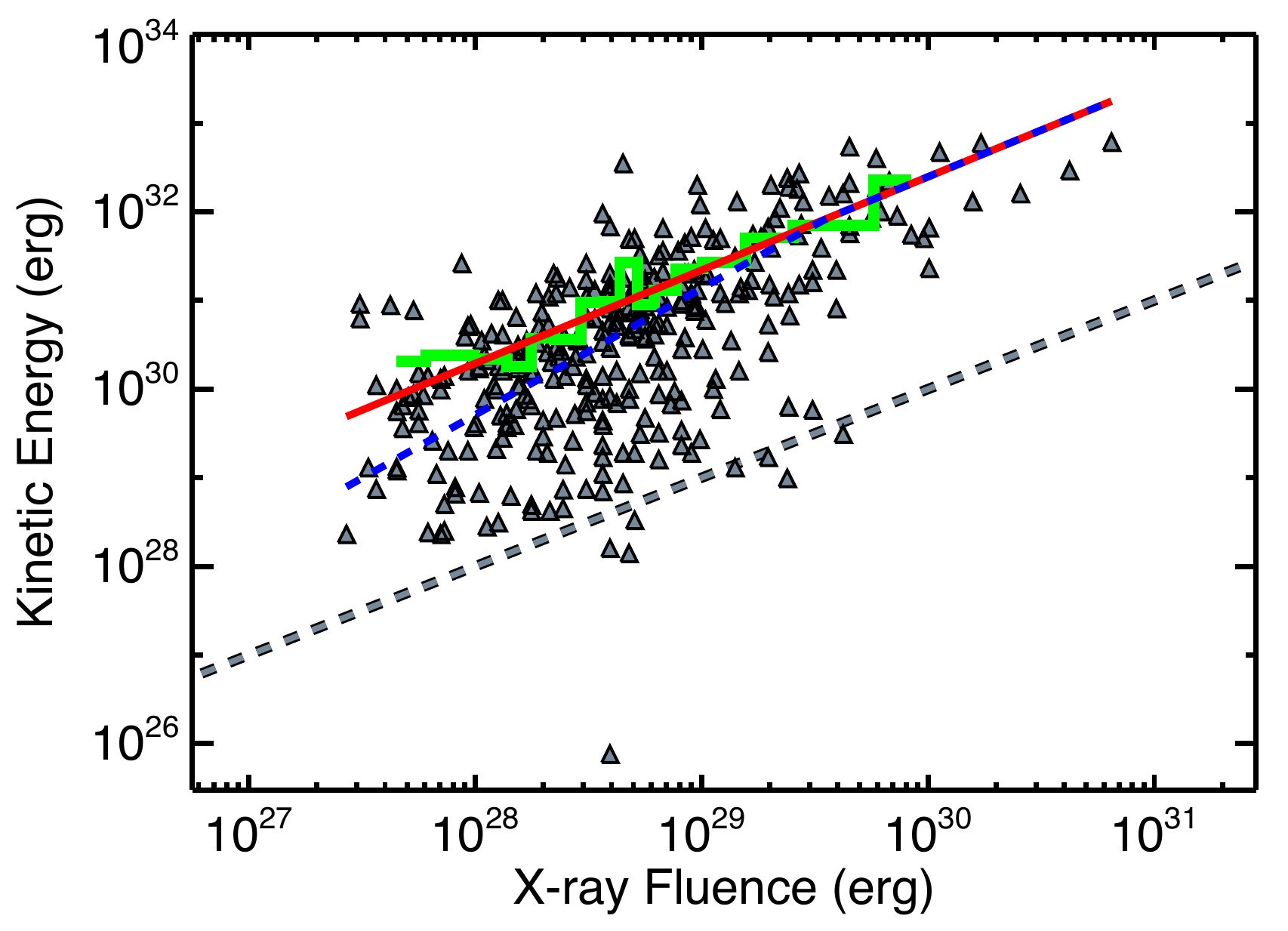}
\caption{CME mass (left) and kinetic energy (right) vs X-ray fluence of the associated flare from the \cite[Yashiro \& Gopalswamy (2009)]{Yashiro.Gopalswamy:09} sample.  The green histograms are the means over twenty data points and the red lines are linear fits to these means.  
The dashed blue lines are the linear fits multiplied by the CME-flare association rate (see text).  In the left panel, the dashed grey line follows a constant ratio of mass loss to GOES X-ray energy loss, expressed as rates, $\dot{M}=10^{-10}(L_X/10^{30}) M_\odot$~yr$^{-1}$.
In the right panel, the dashed grey line represents equivalence of the kinetic and X-ray energies.  The red line in this panel corresponds very closely to a factor of two hundred times the X-ray fluence.  
\label{f:cme_lx}}
\end{figure}

We can also fit the \cite[Yashiro \& Gopalswamy (2009)]{Yashiro.Gopalswamy:09} CME and associated flare data to similar power laws.  The relationships between flare X-ray fluence and CME mass and kinetic energy are illustrated in Figure~\ref{f:cme_lx}, together with our power-law fits, including the flare energy-dependent CME association rate illustrated in Figure~\ref{f:evlac}.  We can combine these power-laws with Eqn.~\ref{e:hudson} and a power-law approximation to the flare-CME association rate (Fig.~\ref{f:evlac}; $f(E)= \zeta E^\delta \,\,\, {\rm for}\,\,\,E \leq 3.5\times 10^{29}$~erg, $f(E)= 1\; {\rm for}\,\, E > 3.5\times 10^{29}$~erg, $\zeta=7.9\times10^{-12}, \,\, \delta=0.37$) 
to obtain the following expressions for the total CME mass and kinetic energy loss rates: 
\begin{equation}
\dot{M}_c=\mu \zeta L_X \left(\frac{2-\alpha}{1+\beta+\delta-\alpha}\right) \left[\frac{E_{max}^{1+\beta+\delta-\alpha}-E_{min}^{1+\beta+\delta-\alpha}}{E_{max}^{2-\alpha}-E_{min}^{2-\alpha}}\right] ,
\label{e:mcme}
\end{equation}
\begin{equation}
\dot{E}_{ke}=\eta \zeta L_X \left(\frac{2-\alpha}{1+\gamma+\delta-\alpha}\right) \left[\frac{E_{max}^{1+\gamma+\delta-\alpha}-E_{min}^{1+\gamma+\delta-\alpha}}{E_{max}^{2-\alpha}-E_{min}^{2-\alpha}}\right].
\label{e:kecme}
\end{equation}
The constants $\mu$, $\beta$, $\eta$ and $\gamma$ are from the power law fits relating the CME mass and kinetic energy to X-ray fluence: $m_c(E)=  \mu E^\beta, \, \mu=10^{-1.5\mp 0.5}, \, \beta=0.59\pm 0.02$; $E_{ke}= \eta E^\gamma, \, \eta=10^{0.81\mp 0.85}, \,\gamma=1.05\pm 0.03$.  For fiducial flare energy limits we have adopted $E_{max}=10^{34}$~erg and $E_{min}=10^{-6}E_{max}$.  The former corresponds to a reasonably large but fairly common flare on an active solar-type star (but is still a thousand times more energetic that large solar flares!).  It turns out that these relationships are fairly insensitive to the exact value of the energy limits and the value of $\alpha$ adopted. 

\begin{figure}
\includegraphics[width=2.6in]{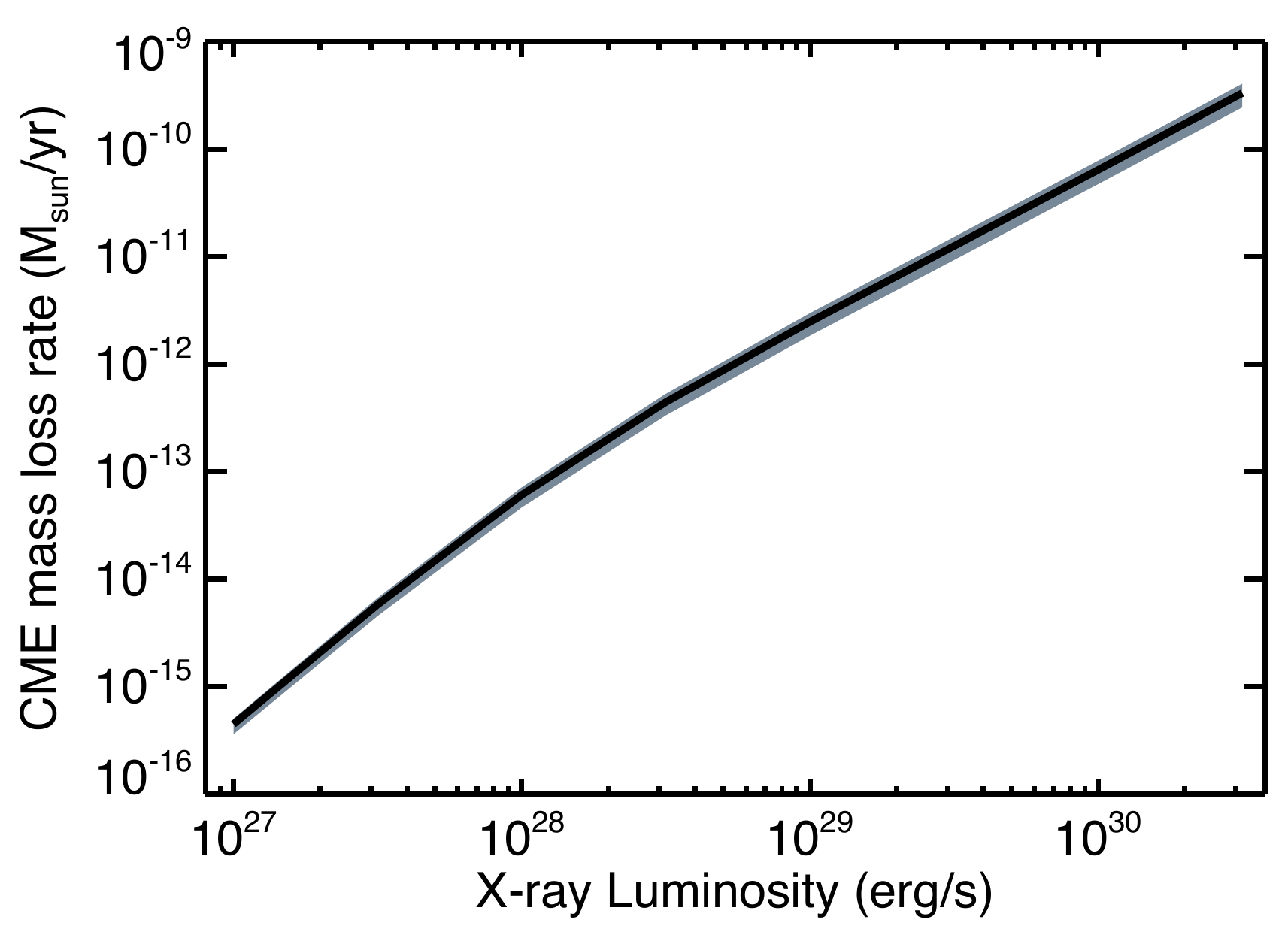}
\includegraphics[width=2.6in]{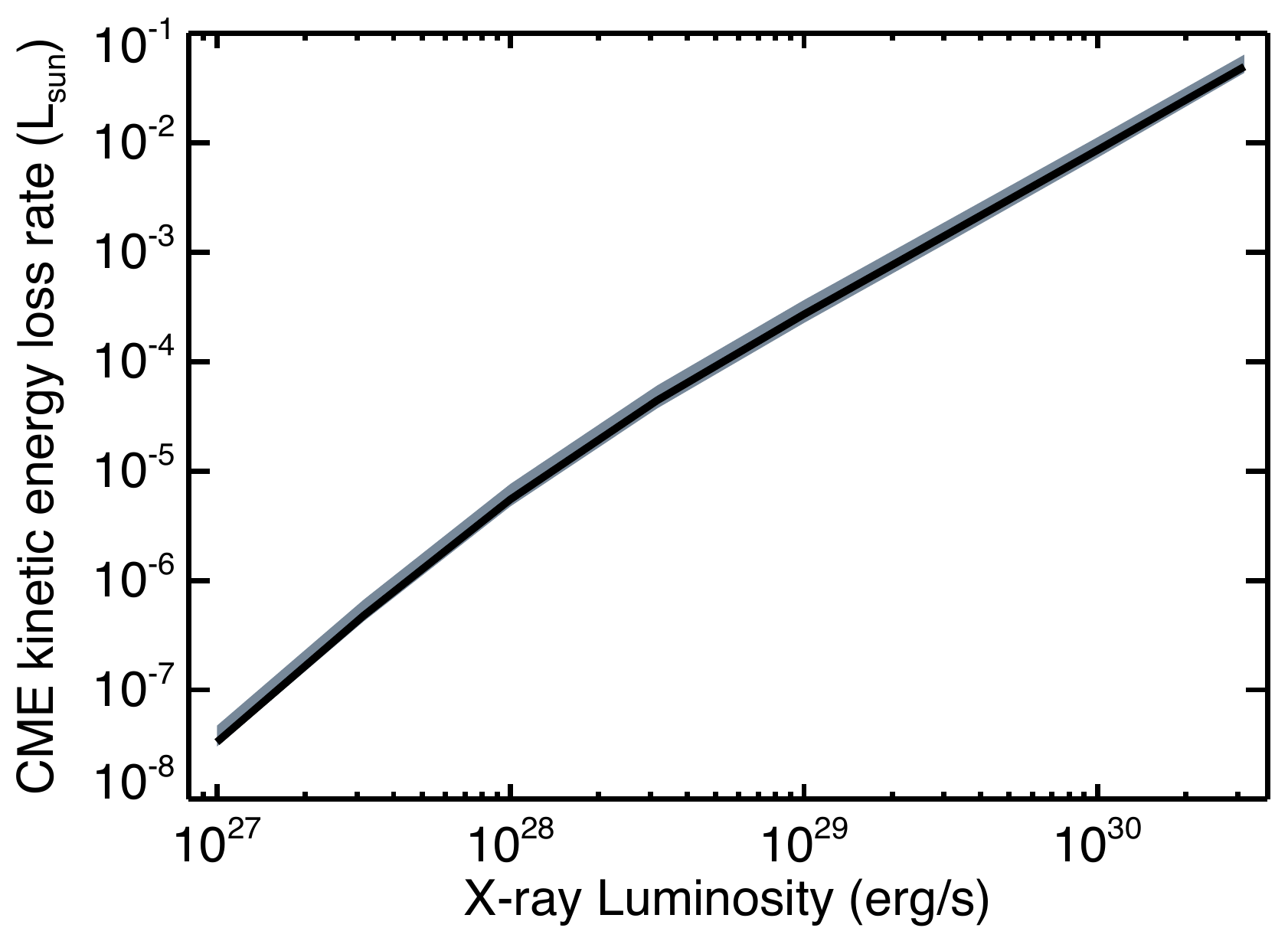}
%\vspace{-0.6in}
%\hspace{-1.5in}\includegraphics[width=0.8in]{duke_orig.png}
\caption{CME mass (left) and kinetic energy (right) loss rates vs.\  broad-band X-ray luminosity. The solid curve represents the power law index $\alpha=2.25$.  The grey shaded areas represent the range in the loss rates corresponding to the power law index range $1.5 \leq \alpha \leq 3.0$.
\label{f:lx_loss}}
\end{figure}

Using an $L_X$-dependent scaling factor to obtain a broad-band stellar $L_X$ from the GOES 1--8~\AA\ band (see \cite[Drake et al.\ 2013]{Drake.etal:13}), we can obtain the CME mass and energy loss rates as a function of broad-band X-ray luminosity.  These are illustrated in Figure~\ref{f:lx_loss}.  Here, the minimum and maximum flare energies were assumed to be $E_{min}=10^{-6}E_{max}$ and $E_{max}=10^4L_X$, though again the results depend only weakly on the exact limits of integration. 

\vspace{-0.1in}

\section{Laws of Physics}
\label{s:physics}

One striking conclusion from Figure~\ref{f:lx_loss} is that, for a solar-like star at saturated activity levels and with $L_X\sim 10^{30}$~erg~s$^{-1}$, the CME mass loss rate is $\dot{M}\sim 5\times 10^{-10}M_\odot$~yr$^{-1}$.   Such a high mass loss rate could be extremely important for understanding a major puzzle in the evolution of Earth and emergence of life: the ``faint young Sun paradox''. \cite[Sagan \& Mullen (1972)]{Sagan.Mullen:72}  pointed out that the lower solar luminosity predicted by stellar evolutionary theory earlier in the history of the solar system implies that for contemporary albedos and atmospheric composition global mean temperatures would have been below the freezing point of seawater until about 2.3~Gyr ago, in contradiction with geological evidence for liquid oceans.  One solution is a more massive Sun at earlier times that was consequently more luminous  (e.g. \cite[Sackmann \& Boothroyd 2003]{Sackmann.Boothroyd:03}).  The CME mass loss rate we have derived, integrated over a few hundred million years when the Sun would have been very active, would appear to be enough to solve the problem (see \cite{Drake.etal:13} for further discussion)!
 
On the down side, such high mass loss rates would produce winds opaque to radio waves and are then at odds with radio detections indicating circumstellar radio transparency (\cite[Lim \& White 1996]{Lim.White:96}).  The scant available estimates of mass loss rates for solar-type stars based on Ly$\alpha$ absorption are also orders of magnitude lower (\cite[Wood et al.\ 2014]{Wood.etal:14}).  While both of these empirical constraints are based on an assumption of steady mass loss, the situation should not be too different for the cumulative effect of a superposition of quasi-continuous CMEs. 

Flogging what would seem to be a severely disadvantaged, if not dead, horse, the corresponding CME kinetic energy requirement approaches $\dot{E}_{ke}\sim 0.1L_\odot$.   In the context of current ideas concerning mass loss and efficiency of magnetic energy dissipation on active late-type stars these values are extremely high!  Surely there is some mistake?  The origin of the high loss rates can be seen straightforwardly in Figure~\ref{f:cme_lx} that shows the vector (grey dashed line) corresponding to a constant ratio of mass loss to GOES X-ray energy loss converted to loss rates, $\dot{M}=10^{-10}(L_X/10^{30}) M_\odot$~yr$^{-1}$, in the upper panel.  Similarly, in the lower panel, the vector corresponding to the equivalence of X-ray and kinetic energy is shown: the mean CME kinetic energy is 200 times that of the radiated GOES band X-ray energy.  For stars whose X-rays are all flares, and are saturated at $L_X/L_{bol}\sim 10^{-3}$, it is straightforward to see we would need $0.1L_{bol}$ to power the CMEs.   

Since the laws of physics have to intervene somewhere (though it often appears the laws of astrophysics must be fundamentally different), there must have been some impropriety committed in extrapolating the solar data to very active stars.  But where has it gone wrong?

%\vspace{-1.5in}
%\hspace{4in}\includegraphics[width=1.2in]{duke_orig.png}
%\vspace{2in}

%\hspace{4.8in}\includegraphics[height=0.3in]{duke_orig.png}
%\vspace{-0.5in}

\vspace{-0.2in}

\section{The nature of CMEs on active stars}

We of course do not yet know much about CMEs on active stars and can only guess how they might be similar or different to solar examples.  We can gain further clues from computer models.  Simulations of the solar wind using numerical MHD models have reached a mature level such that wind conditions can be fairly accurately predicted given a good quality surface magnetic field map (``magnetogram"; see, e.g., \cite[Cohen et al.\ 2007]{Cohen.etal:07}).  While realistic {\it ab initio} simulations of flares and CMEs, beginning with an observed surface magnetic field configuration and evolving this until the events occur, remains difficult, artificial CMEs that look just like those observed can be triggered by inserting an unstable field configuration
%---a \cite[Titov \& D\'emoulin (1999)]{Titov.Demoulin:99} flux rope---
into a time-dependent wind simulation (e.g., \cite[Cohen et al.\ 2010]{Cohen.etal:10}).  We can try the  same thing for the magnetic field configuration of an active star.

\begin{figure}
%\hspace{1.5in}\includegraphics[height=0.5in]{duke_orig.png} \vspace{-0.5in}\\
%\begin{center}
\includegraphics[width=2.6in]{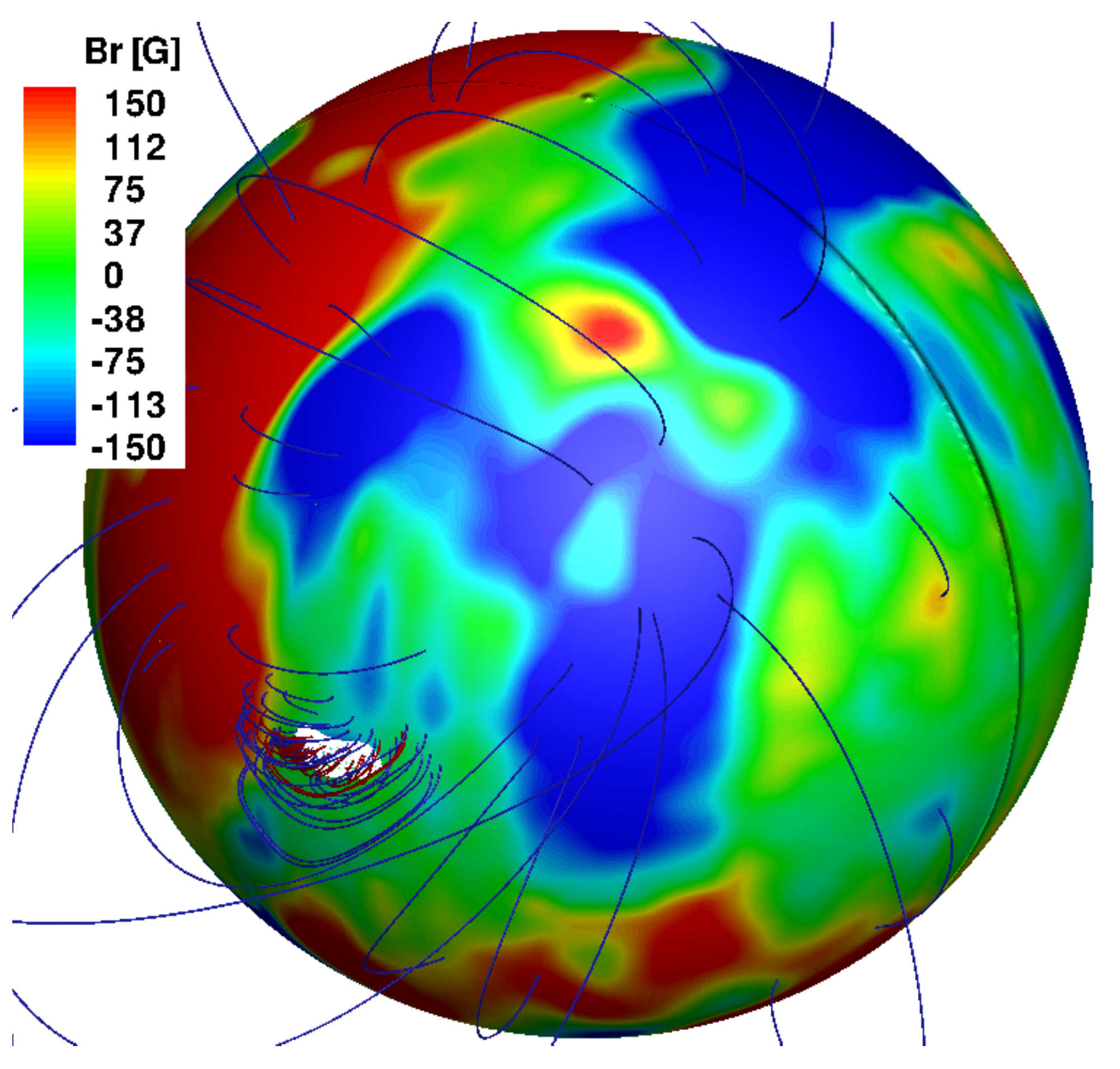}
\includegraphics[width=2.6in]{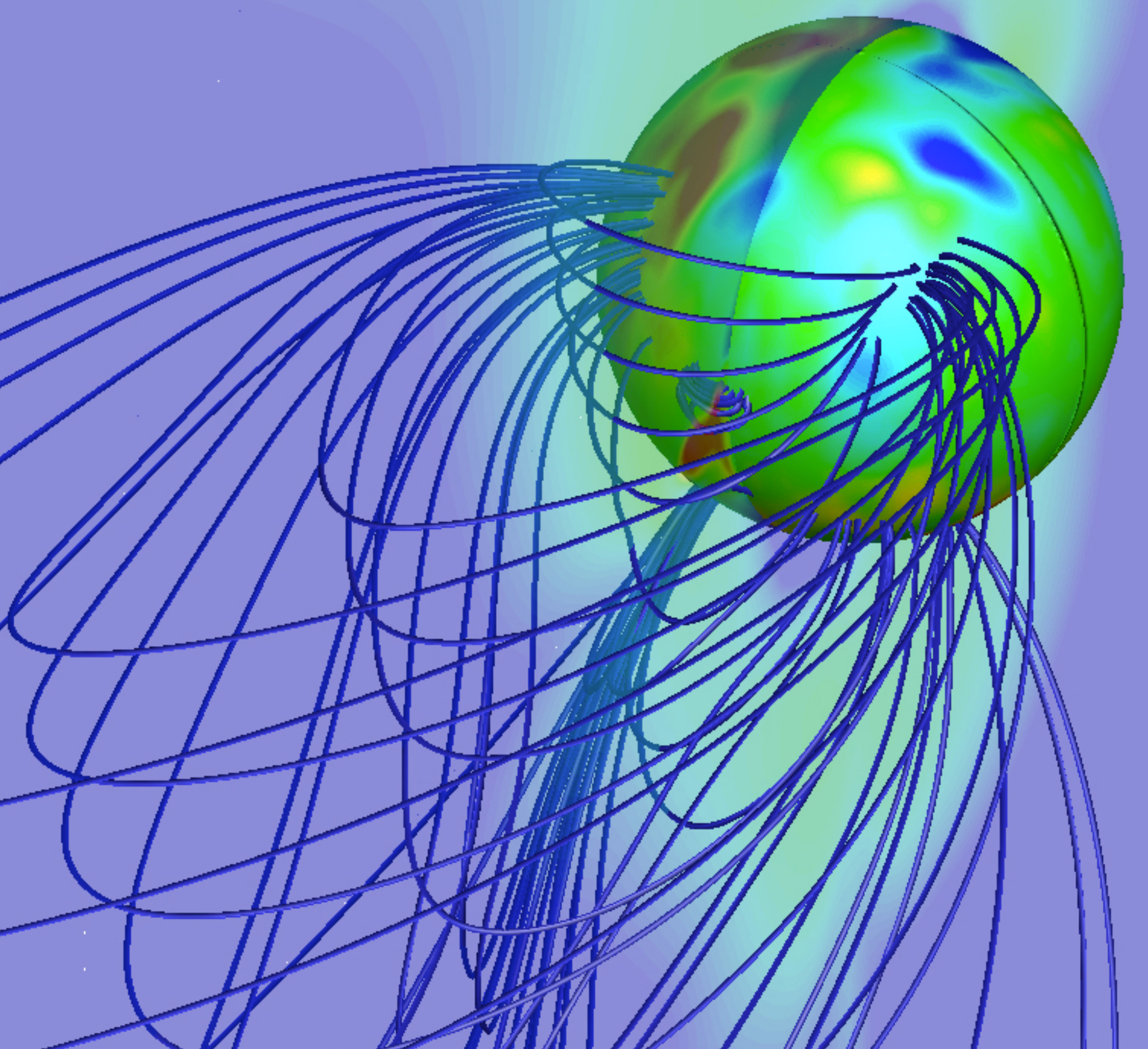}% \vspace{-2.0in}\\
%\hspace{1.5in}\includegraphics[height=0.5in]{duke_orig.png}\vspace{1.5in}\\
%\hspace{-1.2in}
%\end{center}
\caption{A simulation of what would be a large CME on the Sun transposed to the corona of the active K1~dwarf AB~Doradus.  The left panel shows the surface radial magnetic field obtained through Zeeman-Doppler imaging, together with the  artificial Titov-D'emoulin flux rope (a configuration of torodial and poloidal currents that drives the flux-rope unstable) prior to the CME initiation.  The right panel shows the CME close to the peak of its extent within the stellar magnetosphere.  The red material is the CME plasma, which never manages to progress further than a few tenths of a stellar radius.  Overlying magnetic field lines are barely perturbed.
\label{f:cme}}
\vspace{-0.1in}
\end{figure}

We have developed the {\sc BATS-R-US} MHD code used for solar wind, CME and space weather simulations to the case of other stars, using observed Zeeman-Doppler imaging (ZDI) magnetograms as the basis of the coronal model and adopting the same Alfv\'en wave wind driving mechanism (\cite[Oran et al.\ 2013; van der Holst et al.\ 2014]{Oran.etal:13,van_der_Holst.etal:14}).  Figure~\ref{f:cme} illustrates the initial condition for a CME on the young, active K1 dwarf AB~Dor.  The steady-state coronal model and wind is an updated version of that described in \cite[Cohen et al.\ 2010]{Cohen.etal:10}.   A CME is initiated using a \cite[Titov \& D\'emoulin (1999)]{Titov.Demoulin:99} unstable flux rope with a total free energy of $5\times 10^{32}$~erg and a mass of $5\times 10^{16}$~g.  Comparison with Figure~\ref{f:cme_lx} will reveal that such parameters correspond to the largest, most energetic CMEs on the Sun.  

The evolution of the CME with time was followed, one snapshot of which is shown in the right panel of Figure~\ref{f:cme}.  The CME material is the small red splodge between 7 and 8 o'clock, and this was actually the maximal radial extent of its trajectory.  Unlike the situation in a large solar CME, the overlying magnetic field lines remained unbroken, and in fact were barely perturbed by the event.   In this simulation, the Titov-D\'emoulin flux rope was placed in a region between opposite polarities as would be expected.  These regions can have overlying magnetic fields of up to 100 times the few Gauss large-scale fields on the Sun.  The energy of the event was simply too small to facilitate breakout from the strong ambient stellar field.   

Magnetic suppression would appear to be a viable mechanism for attenuating CMEs on active stars.  Of course, the CMEs that will get attenuated are the ``weak" ones---monster CMEs of much greater energy (that are unfortunately numerically difficult to simulate) would be expected to behave like scaled-up solar ones.  However, this suppression mechanism is a promising way to alleviate our mass and energy quandary from \S\ref{s:physics}.  The key lies in the value of $\alpha$ in Eqn.~\ref{e:hudson}: for values of $\alpha < 2$, the preponderance of flare energy lies in the largest events, whereas for $\alpha > 2$ most of the energy is in smaller events.  Our X-ray light curve analyses indicate the latter is the case for active stars---most of the observed X-ray energy is from smaller events.  Further work is needed to understand what the magnetic suppression threshold---in reality it will be a more gradual cut-off---might be for stars of a given activity level. 

There is a recently documented solar analogue of the process we are proposing to avoid the mass-energy CME catastrophe.  Active region NOAA 2192 observed in 2014 October was a particularly large active region and notable for producing 6 X-class flares in a 9 day period.  However, all these flares occurred in the absence of CMEs.  \cite[Thalmann et al.\ (2015)]{Thalmann.etal:15} concluded that an overlying North-South oriented magnetic arcade suppressed CMEs, much like the strong overlying field on AB~Dor in our simulation of Figure~\ref{f:cme}.

\vspace{-0.1in}

%\bibliographystyle{apj}
%\bibliography{cme,batsrus,jjdrake,flares,fys,abdor}

\end{document}